  \documentclass[prl,aps,twocolumn,showpacs]{revtex4}
\usepackage[dvips]{graphicx}
\usepackage{dcolumn}
\newcommand{\be}{\begin{equation}}
\newcommand{\ee}{\end{equation}}
\newcommand{\bea}{\begin{eqnarray}} 
\newcommand{\ba}{\begin{array}}
\newcommand{\eea}{\end{eqnarray}}    
\newcommand{\ea}{\end{array}}

\begin{document}

\title{Nanoparticle networks as chemoselective sensing devices} 

\author{ Natalya A. Zimbovskaya$^{1,2,3},$  Mark R. Pederson$^3,$ Amy S. Blum$^3,$ 
 Banahalli R. Ratna$^3,$ and Reeshemah Allen$^3$}

\affiliation{$^1$Department of Physics and Electronics,
 University of Puerto Rico-Humacao, CUH Station, Humacao, PR 00791,}
\affiliation{$^2$Institute for Functional Nanomaterials, University of Puerto 
Rico, San Juan, PR 00931,}
\affiliation{$^3$Naval Research Laboratory, 4555 Overlook Ave SW, 
Washington, DC 20375}

\begin{abstract}
 We theoretically analyzed transport properties of a 
molecular network constructed of gold nanoparticles linked with 
oligophenylenevinulene (OPV) molecules. We showed that the conductance of such 
system was strongly reduced when trinitrotoluene (TNT)
 became attached to the OPV linkers in the network. 
 The reported results are based on the {\it ab initio} 
electronic structure calculations. These results corroborate and elucidate
experiments which revealed significant drops in the conductance the network while 
the latter was exposed to TNT vapors. The results suggest that the detected
sensitivity of transport characteristics of the considered nanoparticle 
network to TNT  may be used to design a sensing  nanodevice.
\end{abstract}

\maketitle
\date{\today}

\begin{figure}[t]
\begin{center}
\includegraphics[width=5cm,height=8.8cm,angle=-90]{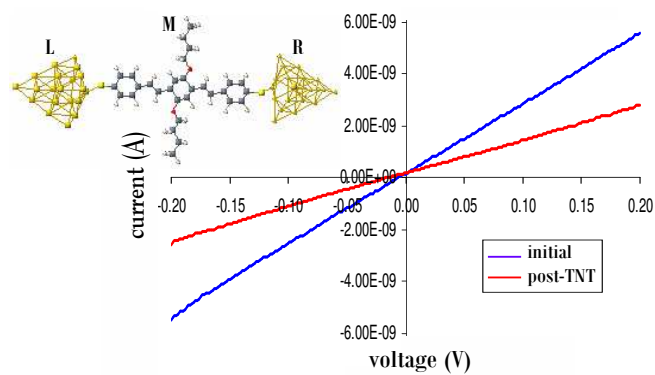}
\caption{ Color online:
The nanoparticle network element 
 and current-voltage characteristics for the network
alone (upper blue line) and the network exposed to TNT vapors (lower red line) obtained in
the experiments.
}
 \label{rateI}
\end{center}\end{figure}

Molecular-scale conductors attract significant interest and attention in both
fundamental and application research during the last two decades \cite{1,2,3,4}.
The unceasing interest of the research community to these conductors originates 
from their actual and potential usefulness in inventing and designing various 
nanodevices. In particular, the integration of nanotechnology with molecular
recognition poses numerous possibilities for creation of novel sensing devices 
\cite{5,6,7,8,9,10}. 
 Significant efforts are being applied to study
electron transport in networks made out of nanoparticles connected by linker
molecules \cite{11,12,13,14,15,16}. It is both interesting and important to
investigate their potential in designing molecule recognizing sensors.

   The present work is motivated by recent experiments  carried out 
to explore transport properties of a three-dimensional
molecular network consisting of gold nanoparticles linked by 
oligophenylenevinulene (OPV) molecules. The molecular network was
built using a cowpea mosaic virus (CPMV) as a scaffold.   
 The CPMV is a particle of an icosahedral geometry made out of sixty copies of a protein subunit whose average diameter is about $ 60 nm , $ as was determined by $X$-ray crystallographical analysis \cite{17}. To use the CPMV  as a scaffold for the molecular network it was genetically engineered in order to provide cystein residues at selected positions. Gold nanoparticles with diameters $ \sim 5 nm$ were bound to these residues. Due to the icosahedral symmetry of the virus the residues and attached gold particles are arranged into complex three dimensional structure of the same symmetry. The latter determines interparticle distances. The CPMVs decorated with gold nanoparticles were exposed to OPV molecules which linked the gold particles producing conductive network on the viruses.


To measure the conductance of a network built on a single virus the latter was assembled between proximal probe electrodes. The electrodes were made out of pair of $100 nm$ wide gold leads patterned on a silicon wafer. After measuring the baseline conduction, the entire chip was exposed to the target, and the transport measurements were repeated after binding the target. Actually the test chip included several pairs of electrodes which allowed statistical analysis of the results.
  Strong changes in the network 
conduction were observed when the latter was exposed to trinitrotoluene
(TNT) vapors. The exposure resulted in a significant decrease in the conduction
(Fig. 1). This conduction decrease occurred at rather small values 
of the bias voltage applied across the network $(-0.2V \div 0.2V).$ The
experimentally disclosed sensitivity of the above described molecular network 
to TNT makes it potentially useful to design a TNT sensing nanodevice. 

To theoretically analyze and explain these experimental results we computed 
eigenenergies and corresponding wave functions for the molecular network
element including two gold nanoparticles connected by the OPV molecule both
with and without attached TNT. 
 We did perform full 
self-consistent computations using the NRLMOL software package \cite{18}. In 
addition to self-consistently solving the Kohn-Sham equations we have optimized 
the geometries of the examined molecules. Once we arrived at the resulting
Hamiltonian matrices we proceeded using Lowden's method \cite{19} of symmetrical 
orthogonormalization to build up atom centered Wannier-like wave functions 
\cite{20} starting from the nonorthogonal gaussian orbitals. Using these as a 
new basis functions we reconstructed the Hamiltonian  matrices. Now, the matrices 
became nearly diagonal enabling us to separate out three diagonalized blocks 
corresponding to the left and right gold nanoparticles $(L,R)$ and the
molecule $(M)$ in between. Small off-diagonal elements in the resulting 
Hamiltonian matrices between the $L-M $ and $M-R $ blocks describe the coupling
of the molecular linker to the gold nanoclusters in the considered network.

The final matrices used in the following transport calculations are obtained 
separating out those parts of the Hamiltonian matrices which could be placed
in the window  around the equilibrium Fermi energies $E_F.$ The window
 width is determined by the value of the bias voltage. So, the full 
Hamiltonian matrices are replaced by these reduced matrices of relatively 
small dimensionalities. This procedure is justified for only those states whose
 eigenenergies are rather close to the Fermi energy could actually contribute to the
electron transport. 

 In our  transport calculations we employ the effective 
Hamiltonian for the molecule linking two adjacent metal nanoparticles. The
latter is written in the usual form (see e.g. Refs. \cite{21,22}):
   \be
 H_{eff} = H_M + H_L + H_R.
    \label{1}\ee
   Here, the term $H_M $ corresponds to the molecule, and $ H_{L,R} $ 
describe the coupling of the molecule to the gold nanoclusters. Omitting 
for a while the last two terms and keeping in mind that the orbitals included
into the basis set are orthogonormalized, we may introduce the retarded
Green's function for the single molecule $ G_0^R (E) $
 \cite{2}. The latter is defined by the matrix equation:
  \be
 \big[(E + i\eta)\hat I - H_M \big] \hat G_0^R (E) = \hat I
   \label{2}   \ee
   where  $\hat I $ is the identity matrix. The parameter $\eta $
describes the broadening of the molecule energy levels due to random motions
of the surroundings at finite temperature $ T $ and takes on values of the
order of thermal energy $ kT. $

To proceed we employ the Dyson equation relating the  retarded Green's 
function of the molecule coupled to the metal nanoclusters $\hat G^R(E) $ to
the Green's function of a single molecule. The equation reads \cite{23}:
   \be
   \hat G^R(E) = \big[\hat A(E) - \hat\Sigma(E)\big]^{-1}.   \label{3}
    \ee
  Within the chosen basis set of the wavefunctions the matrix 
$\hat A(E) = \big[G_0^R  (E) \big]^{-1}$ is a diagonal matrix
   \be 
 A_{ij} (E) = (E + i\eta - E_i)\delta_{ij}   \label{4}
  \ee
  where $E_i$ are the eigenenergies of the molecule.

In the Eq. (\ref{3}), the self-energy term $\hat\Sigma(E)$ characterizes the
coupling of the molecule to the left and right nanoclusters:
  \be
 \hat\Sigma(E) = \hat\Sigma_L(E) + \hat\Sigma_R(E).   \label{5}
  \ee
  
The matrix elements of self-energy corrections have the form \cite{1}:
   \be
 (\Sigma_\beta)_{i,j} = \sum_k
\frac{W_{ik,\beta}^*W_{kj,\beta}}{E-\epsilon_{k,\beta} - i\sigma_{k,\beta}} .
   \label{6}  \ee
  Here, $\beta \in L,R, \ W_{ik,\beta}$ are, respectively, the coupling 
strengths between $i$-th molecule state and $``k" $-th state on the 
left/right  metallic cluster, $\epsilon_{k,\beta} $ are energy levels 
of the corresponding nanoclusters, and the parameters $\sigma_{k,\beta}$ 
describe the thermal broadening of the electron levels at the clusters.
The summation over $k $ in the Eq. (\ref{6}) 
is carried out over the states of the left/right cluster.

When the bias voltage $ V $ is applied across the system shown in the Fig. 1, this causes charge redistribution, and subsequent changes in the energies $E_i$ and $\epsilon_{k,\beta}.$ In consequence, the matrix elements $ A_{ij} $ and $\Sigma_{ij} $ values vary as $V $ changes. This affects 
 the electron transmission function $T $  given by the expression:
   \be
T = Tr \big\{\hat\Gamma^L\hat G^R \hat\Gamma^R\hat G^A \big\} \label{7}
  \ee
  where $\hat\Gamma^{L,R} =- 2\mbox{Im}\hat\Sigma_{L,R},$ and $G^A $ is the 
advanced Green's function of the molecule $\big(\hat G^A = (G^R)^\dag\big)$. 
 Nevertheless, our computations showed that at low bias voltage values $(|V|
< 0.25V) $ its effect on the electron transmission is weak and may be neglected.
Therefore, in further analysis we use the expression for the electron 
transmission function computed at $ V=0.$ 
 We remark,  that the
self-energy parts given by Eq. (\ref{6}) depend on the values 
of $\epsilon_{k,\beta}$ representing  electron structures 
of the gold nanoclusters, and on temperature inserted in the
terms $\sigma_{k,\beta}$.
 So, the changes in the electron structure of the clusters may affect 
the transmission as well as the changes in the electron structure of the 
linking molecule. The latter is included in the expressions for the Green's 
functions $G^{R,A}.$

Now, we employ the standard expression for the electron tunneling current
flowing through the molecule \cite{2}:
  \be
 I = \frac{e}{\pi\hbar} \int_{-\infty}^\infty dE T(E)
\big[f(E - \mu_L) - f(E - \mu_R)\big].   \label{8}
  \ee
 Here, $e$ is the electron charge,  $\hbar$ is the Planck's constant, $T(E)$ 
is the electron transmission given by Eq. (\ref{7}), $f(E)$ is the Fermi
distribution function for the energy $E.$ Chemical potentials $ \mu_{L,R}$ 
are attributed to the left/right  nanoparticles, and they are shifted 
from the equilibrium Fermi energy $E_F$ due to the bias voltage $V$ applied 
across the system:
   \be
 \mu_L = E_F + (1 -\nu) eV, \qquad \mu_R = E_F + \nu eV   \label{9}
  \ee
  where $ \nu $ is the division parameter which shows how the voltage $V$
is distributed between the  nanoclusters.

\begin{figure}[t] 
\begin{center}
\includegraphics[width=4.8cm,height=9.2cm,angle=-90]{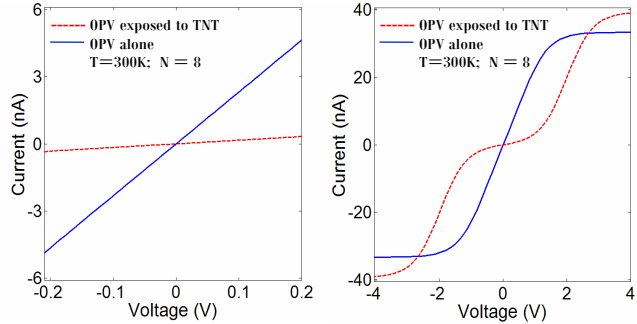}
\caption{ Color online:
Current-voltage characteristics calculated for a chain of nine gold nanoparticles
linked by OPVs alone (solid lines) and by OPV-TNT complexes (dashed
lines). 
 The number of linkers $N=8. $
}
 \label{rateI}
\end{center}\end{figure}

We applied  Eqs. (\ref{7})-(\ref{9}) compute to the tunneling electron current through a
part of the experimentally studied network. The chosen part included 
eight pieces connected in series, each piece consisting of two gold
nanoparticles linked by a molecule. 
 The number of pieces roughly corresponds to the number of gold nanoparticles set along the half of the CPMV circumference. 
 We assumed the bias voltage to be applied
across the whole chain, so the voltage  across two adjacent gold
clusters appeared to be smaller than the net voltage $V.$ Supposing that all
molecular links in the chain are the OPV molecules and none of these are
loaded with TNT,
 we did obtain the current-voltage 
characteristics shown as solid lines in the Fig. 2. At low bias voltage the
characteristic
is an ohmic curve which strongly resembles the experimental $I-V$ curve (see
Fig. 1).  
 Both experimental and calculated $I-V$ curves are symmetrical, and the computed current takes on values close to those observed at the same bias voltage, as follows from comparison of the figures $1$ and $2$ (left panel). 
 When TNT molecules are attached to all OPVs included in 
the chain, our calculations result in the critical reduction of the conductance
at low bias voltage as presented in the Fig. 2. This agrees with the
experimental results.

 We remark, that computed drop in conductance significantly exceeds that observed in the experiments. The mismatch could be reduced if we assume that TNT molecules are attached to some OPV linkers in the network not to everyone. Within the chosen model we can show that the decrease in the number of TNTs attached to the OPV linkers brings enhancement in the conductance of the chain which manifests itself as the increase in the slope of the relevant $I -V$ characteristics at low bias. The corresponding lines  
would be positioned between the dashed and the solid lines in the left panel of the figure $2,$ and the smaller is the number of TNT-OPV complexes the closer is the $I-V$ characteristics to the solid line. The latter corresponds to the chain of gold nanoparticles linked by OPVs alone.

 At higher voltage the conductance of the OPV-TNT molecular
complexes enhances and becomes greater than the conductance of single OPV
molecules.
  As follows from our calculations, 
  at the energies close to $E_F $
self-energy terms $ \Sigma_{L,R}$ accept smaller values for joint molecular 
linkers than for single OPVs connecting gold nanoparticles  in the network.
The current between the nanoparticlies is defined by Eq. (\ref{8}), and at low
values of the applied voltage energies close to $E_F$ give the predominating
contribution to the relevant integral. So, the reduction in $\Sigma_{l,R} $
 results in the reduction of the conductance at low bias voltage.

Finally, in the present work we studied transport properties of the network 
consisting of gold nanoparticles linked by OPV molecules. These studies were
motivated by the experimental results which revealed strong sensitivity of
this system to TNT vapors. We carried out {\it ab initio} electronic structure 
calculations for the network element including two gold nanoclusters connected
by the OPV molecule alone and by  OPV-TNT  
 molecules.  
 Then we applied the results to calculate electron 
current through a chain made out of gold nanoclusters connected by molecular
links. We remark that the simplified structure of the network adopted here, 
prevented us form quantitative comparison of the computed
results with those experimentally obtained  
 for the realistic three-dimensional network of rather complicated 
structure. Nevertheless, we showed that at low bias voltage the conductance of
the considered chain strongly decreases 
 when TNT 
molecule becomes coupled to OPV linkers.
This agrees with the experimental results.
  In whole, the present results reasonably explain the observed sensitivity of the above
described nanoparticle networks to TNT. 
  Also, they elucidate the potential of such networks in designing chemoselective sensing nanodevices.

{\it Acknowledgments:} We thank  G. M. Zimbovsky for help 
with the manuscript. NZ acknowledges support from the ASEE and ONR Summer Faculty Research program.


\begin{thebibliography}{99}

\bibitem{1} R. Friend and M. A. Reeds (Eds),  {\it Physics of electronic 

transport in single atoms, molecules and related nanostructures}, 
Nanotechnology {\bf 15}, S433 (2004).

\bibitem{2} S. Datta, {\it Quantum Transport: Atom to Transistor}
 (Cambridge University Press, 2006).

\bibitem{3} M. A. Reeds and T. Lee (Eds), {\it Molecular nanoelectronics} 
(American Scientific Publishers, 2003).

\bibitem{4}  G. Cuniberti, G. Fagas, and K. Richter  (Eds), {\it Introduction 
to Molecular Electronics} (Springer, Berlin, 2005).


\bibitem{5} Y. Cui, Q. Q. Wei, H. K. Park, and C. M. Lieber, Science {\bf 293},
1289 (2001).

\bibitem{6}  Z. Li, Y. Chen, X. Li, T. I. Kamins, K. Nauka, and R. S. Williams,
Nano Lett. {\bf 4}, 245 (2004).

\bibitem{7} Z. Y. Fan and J. G. Lu, Appl. Phys. Lett. {\bf 86}, 123510 (2005).

\bibitem{8} B. L. Allen, P. D. Kichambare, and A. Star, Adv. Mater. {\bf 19}, 
1439 (2007).

\bibitem{9} B. Li, L. Shang, M.S. Marcus, T. L. Clare, E. Perkins, and R. L.
Hamers,  {\bf small 4}, 795 (2008).

\bibitem{10} Q. Wan, E. Dattoli, and W. Lu, {\bf small 4}, 451 (2008).

\bibitem{11} H. D. Sikes, J. F. Smalley, S. P. Dudek, A. R. Cook, M. D. Newton, 
C. E. D. Chidsey, and S. W. Feldberg, Science {\bf 291}, 1519 (2001).

\bibitem{12} A. Blum, C. M. Soto, C. D. Wilson, T. L. Brower, S. K. Pollack, 
T. L. Schull, A. Chatterji, T. Lin, J. E. Johnson, C. Amsinck, P. Franzon,
R. Shashidhar, and B. R. Ratna, {\bf small 1}, 702 (2005).

\bibitem{13} J. G. Kushmerick, D. B. Holt, S. K. Pollack, M. A. Ratner, 
J. C. Yang, T. L. Schull, J. Naciri, M. H. Moore, R. Shashidhar, J. Am. Chem. 
Soc. {\bf 124}, 10654 (2002).

\bibitem{14} A. S. Blum, J. C. Yang, R. Shashidhar, and B. R. Ratna, Appl. 
Phys. lett. {\bf 82}, 3322 (2003).

\bibitem{15} A. Zaber-Khosousi, P. E. Trudeau, Y. Suganuma, and A. Dhirani,
Phys. Rev. Lett. {\bf 96}, 156403 (2006).

\bibitem{16}  P. Nickels, M. M. Matsushita, M. Minamoto, S. Komiyama, and
T. Sugawara, {\bf small 4}, 471 (2008).

\bibitem{17} T. Lin, Z. XChen, R. Usha, C. V. Stanffacher, J. Dai, T. Schmidt, and J. E. Johnson, Virology, {\bf 265}, 20 (1999).

\bibitem{18} M. R. Pederson, D. V. Porezag, J. Kortis, and D. C. Patton, 
Phys. Status Solidi B {\bf 217}, 197 (2000).

\bibitem{19} P. O. Lowden, J. Chem Phys. {\bf 18}, 365 (1950).

\bibitem{20} M. R. Pederson and C. C. Lin, Phys. Rev. B {\bf 35}, 2273 (1985).

\bibitem{21} Y. Xue, S. Datta, and M. A. Ratner, J. Chem. Phys. {\bf 115}, 
4292 (2001).

\bibitem{22} M. Yu. Galperin, M. A. Ratner, and A. Nitzan, J. Chem. Phys. 
{\bf 121}, 11965 (2004).

\bibitem{23} M. Yu. Galperin, A. Nitzan, and M. A. Ratner, Phys. Rev. Lett. 
{\bf 86}, 166803 (2006). 

\bibitem{24} N. A. Zimbovskaya and G. Gumbs, Appl. Phys. Lett. {\bf 81} 
1518 (2002).



\end{thebibliography}
\end{document}